# Covariant Space-Time Line Elements in the Friedmann–Lemaitre–Robertson–Walker Geometry

David Escors * and Grazyna Kochan

Fundación Miguel-Servet, Public University of Navarre, 31008 Pamplona, Spain; grazyna.kochan@navarra.es
* Correspondence: descorsm@navarra.es

**Abstract:** Most quantum gravity theories quantize space-time on the order of Planck length $(\ell_p)$. Some of these theories, such as loop quantum gravity (LQG), predict that this discreetness could be manifested through Lorentz invariance violations (LIV) over travelling particles at astronomical length distances. However, reports on LIV are controversial, and space discreetness could still be compatible with Lorentz invariance. Here, it is tested whether space quantization on the order of Planck length could still be compatible with Lorentz invariance through the application of a covariant geometric uncertainty principle (GeUP) as a constraint over geodesics in FRW geometries. Space-time line elements compatible with the uncertainty principle are calculated for a homogeneous, isotropic expanding Universe represented by the Friedmann–Lemaitre–Robertson–Walker solution to General Relativity (FLRW or FRW metric). A generic expression for the quadratic proper space-time line element is derived, proportional to Planck length-squared, and dependent on two contributions. The first is associated to the energy–time uncertainty, and the second depends on the Hubble function. The results are in agreement with space-time quantization on the expected length orders, according to quantum gravity theories, and within experimental constraints on putative LIV.

## 1. Introduction

General relativity (GR) is a background-independent geometric theory for gravitation in which the space-time metric is the dynamical variable [1,2]. The solutions to Einstein's field equations correspond to space-time metrics defined by pseudo-Riemannian metric tensors ($g_{\mu\nu}$). In GR, particle trajectories follow geodesics in the geometries defined by such metric tensors. As the momentum/position phase space is continuous in classical GR [2], both momentum and position can be simultaneously known with absolute certainty within the geodesic trajectories. As a consequence, particle geodesics can be defined with absolute precision. This clashes with quantum mechanics, in which the momentum/position phase space is quantized. As a consequence, the measurement of position introduces uncertainty in momentum and vice versa. Likewise, an uncertainty relationship also exists between energy fluctuations and time intervals. These two uncertainty relationships constitute the classical Heinsenberg´s uncertainty principle inequalities of quantum mechanics [3,4]:

$$|\Delta p \Delta x| \geq \frac{\hbar}{2} \, , |\Delta E \Delta t| \geq \frac{\hbar}{2}. \tag{1}$$

where $\Delta p$ and $\Delta x$ represent the change in the magnitude of momentum and position, respectively; $\Delta E$ and $\Delta t$ represent the change in magnitude of energy and time, respectively; $\hbar$ is the reduced Planck constant.

These inequalities reflect the quantization of the phase space in units of $\hbar$. These inequalities indicate that the subjacent space-time geometry must also be fundamentally dis-crete. Attempts to quantize the space-time geometry by several methods have led to quantum gravity theories such as Loop Quantum Gravity, or LQG [5,6]. From the process of space quantization, current quantum gravity theories rely on a minimum length for their formulation, which is proportional to Planck length ($\ell_p$). For example, the quanta of area and volume operators in LQG are proportional to $\ell_p^2$ and $\ell_p^3$, respectively [7–9]. Likewise, other quantum gravity theories, such as string theory, also introduce $\ell_p$ as a fundamental length element for particles [10–12]. Most quantum gravity theories predict phenomena such as in vacuo dispersion of photons and neutrinos, and deviations of photon polarization over astronomical distances caused by Lorentz invariance violations (LIV) [13–18]. These phenomena are predicted to arise from current quantum gravity theories if indeed space is discreet.

The process of spatial quantization, or the establishment of a fundamental length associated to particles, alter the classical uncertainty principle, leading to formulations such as the Generalized Uncertainty Principle, or GUP [19–21]:

$$|\Delta p \Delta x| \geq \frac{\hbar}{2} + \frac{\hbar c^2}{m^2 A^2 \delta s^2} \Delta p^2. \tag{2}$$

This expression includes a correction dependent on the particle mass, *m*, the proper acceleration, *A*, the speed of light, *c*, and the quadratic form of the space-time line element $\delta s^2$.

As the space-time metric in GR is shaped by energy-momentum densities through the energy-momentum tensor, vacuum energy and momentum fluctuations from the uncertainty principle should perturb the space-time geometry [19,22]. Indeed, very early on it was assumed that in the context of a quantum description of gravity, quantum fluctuations caused by Heisenberg´s principle play a major role [23]. In semi-classical descriptions of quantum gravity, a putative metric tensor operator, $\hat{g}_{\mu\nu}$, is decomposed in the classical pseudo-Riemannian metric tensor, $g_{\mu\nu}$, and a fluctuating tensor operator of quantum origin, $\delta \hat{g}_{\mu\nu}$, which introduces a differential perturbation [24]:

$$\hat{g}_{\mu\nu} = g_{\mu\nu} + \delta \hat{g}_{\mu\nu}. \tag{3}$$

The indices, denoted by Greek letters take on the values 0, 1, 2 and 3, defining the temporal and spatial components in standard relativistic tensor notation. The expectation value of the perturbation is then identified with a quantum-associated classical tensor $T_{\mu\nu}$:

$$\langle \delta \hat{g}_{\mu\nu} \rangle \equiv T_{\mu\nu}. \tag{4}$$

A more direct relationship between the uncertainty principle and alterations to the metric can be formulated with quantum mechanics commutators used in string theories and quantum topology [19]. A momentum-position commutator is thus associated to the Minkowski metric tensor, that is then generalized to curved space-time through a pseudo-Riemannian metric tensor:

$$[P^\mu, X^\nu] = -i\hbar \eta^{\mu\nu}, [P^\mu, X^\nu] = -i\hbar g^{\mu\nu}. \tag{5}$$

where $P^\mu, X^\nu$ stand for the components of momentum and position 4-vector operators, and $\eta^{\mu\nu}, g^{\mu\nu}$ represent the contravariant Minkowski and pseudo-Riemannian metric tensors, respectively.

Independent of the specific model for quantum gravity, uncertainty fluctuations introduce a perturbation in the metric that is unrelated to classical gravitation, but can be otherwise related to a minimal length for the space-time line element [19,22,25–27]. Likewise, the description of quantum mechanics phenomenology must be compatible with an existing metric background. Perturbation theory can hence be applied to achieve this within a quantum mechanics formalism as described in [28]. However, the use of a fixed length for the line element clashes with classical relativity. Indeed, the LQG minimal length could be considered a "free parameter" [29], complying with Lorentz co-variance [21]. The combination of the uncertainty principle with the requirement for a minimum length led to the development of the Generalized Uncertainty Principle (GUP) in various ways [30–33]. Possibly the most widely used is based on modifications of the position-momentum commutator as extensively described in [27,34,35]. The imposition of Lorentz co-variance has led to corrections to the canonical GUP momentum-position commutator for Minkowski space as shown in [21]:

$$[P^\mu, X^\nu] = -i\hbar\eta^{\mu\nu} - i\hbar A\eta^{\mu\nu} - i\hbar\, B(P^\mu, P^\nu). \tag{6}$$

with A and B being functions of momentum.

Several other mathematical approaches have been undertaken to make compatible the relativistic Lorentz invariance with the uncertainty principle. One of such strategies has been developed within the ADM formalism by adding small position uncertainties, although this approach did break covariance [36]. The imposition of Lorentz invariance can also be problematic in quantum field theories due to the difficulties arising from calculating amplitudes with Minkowski signatures. One way to overcome this difficulty was recently shown by calculating first the amplitudes in Euclidean space, followed by deriving analytic solutions to the Minkowski signature [37].

We recently reformulated the classical uncertainty principle in a relativistic covariant form in terms of the proper space-time line element ($\tau^2$) and Planck length, $\ell_p$ [38]. This reformulation allows its application as a mathematical constraint over GR geodesics. The differential quadratic proper space-time line element is then defined as a function of Planck length through a geodesic-derived scalar, $G_{\text{geo}}$:

$$|G_{\text{geo}}\, d\tau^2| \geq (1+\gamma)\, \ell_p^2. \tag{7}$$

where $d\tau$ is the proper space-time line element, and the gamma factor $\gamma$ and Geodesic scalar are defined in units of $c$ set to 1:

$$\gamma = \frac{dt}{d\tau} \equiv \frac{E}{m},$$
$$G_{\text{geo}} \equiv 2Gm\left|U_0 \Gamma^0_{\alpha\beta} U^\alpha U^\beta\right| + 2Gm\left|U_j \Gamma^j_{\alpha\beta} U^\alpha U^\beta\right|. \tag{8}$$

where $E$ corresponds to the total energy of the particle; $m$ corresponds to its mass; $\Gamma^\mu_{\alpha\beta}$ corresponds to Christoffel symbols calculated from the pseudo-Riemannian metric tensor; G is the universal gravitational constant; $U_\alpha$, $U^\alpha$ are covariant and contravariant components of proper velocity. The indices denoted by j take on the values 1, 2 and 3, defining the spatial components in standard relativistic notation. This formulation sets a length limit for the quadratic proper space-time line element proportional to $\ell_p^2$, which can be applied to GR geodesics as a constraint [38].

The exact solution to Einstein´s field equations for a homogeneous, isometric Universe that expands following Hubble´s law corresponds to the FRW (or FLRW) metric [39,40]. This solution represents a first approximation to the standard model of Cosmology. Its line element is represented with a (− + + +) Lorentzian metric signature:

$$d\tau^2 = -dt^2 + a^2\frac{dR^2}{1-KR^2} + a^2R^2 d\theta^2 + a^2R^2\sin^2\theta d\varphi^2. \tag{9}$$

where $t$ corresponds to the temporal coordinate and $R$, $\theta$ and $\varphi$ to dimensionless co-moving polar coordinates. The time-dependent universal scale factor, $a$, provides dimensions of length to the co-moving coordinates, and determines the physical size of the Universe; The curvature constant, $K$, takes on values of 1, 0 or −1, depending on the model of the expanding Universe; closed-spherical, flat or open-hyperbolic, respectively. For the space-time quadratic line element, $d\tau^2$ notation is chosen instead of $ds^2$ as these are equivalent in units of c set to 1. Although the Universe presents a very small positive curvature, for practical purposes it can be discarded. Indeed current measurements of cosmological parameters are in agreement with a spatially-flat cosmology [41]. Hence, $K$ set to zero could be justified to describe the current state of the Universe.

In principle, it could be considered that the lengths for the space-time line element in the FRW metric depend on two factors: first, energy–time quantum fluctuations, and second, the expansion rate of the Universe as defined by Hubble's function, $H$, calculated from the scale factor:

$$H = \frac{\dot{a}}{a} \ ; \ \dot{a} \equiv \frac{da}{dt} \tag{10}$$

Here, GeUP is applied as a constraint to the FRW solution to define Lorentz invariant length restrictions on the space-time line element. The results from the calculations are then discussed in the context of the predicted length elements in quantum gravity theories, and the experimental restrictions upon Lorentz invariance violations calculated from experimental observations.

## 2. Results

### 2.1. Space-Time Line Element for a Particle at Rest in the FRW Metric

Unless otherwise stated, the equations will be expressed in units of c set to 1. Standard relativistic tensor notation will used, with temporal and polar spatial coordinates designated as: $(X^0, t); (X^1, R); (X^2, \theta); (X^3, \varphi)$. The most general FRW solution is given by Equation (9). A particle at rest implies the following statements on proper velocities:

$$U^1 = U^2 = U^3 = 0, \ U^0 U_0 = -1. \tag{11}$$

To apply GeUP from Inequality (7), the geodesic scalar is first calculated:

$$G_{geo} = 2Gm \left| U_0 \Gamma^0_{00} U^0 U^0 \right|. \tag{12}$$

The only Christoffel connector that participates in the calculation is $\Gamma^0_{00}$, which is zero in the FRW metric. Hence, the geodesic scalar is also 0, leading to a contradiction in Inequality (7) unless Plank length is considered zero in the non-quantum limit:

$$0 \geq (1 + \gamma)\ell_p^2. \tag{13}$$

Therefore, the classical FRW metric is incompatible with Inequality (7) unless a t-dependent differential perturbation function, $\varepsilon$, is introduced in the $g_{00}$ component of the metric, following a similar approach by semi-classical quantum gravity and developed for Minkowski space in [38]:

$$g_{00} = -1 - \varepsilon(t). \tag{14}$$

The introduction of this differential perturbation in the FRW metric leads to a compatible solution with the uncertainty principle without modifying the overall solution:

$$d\tau^2 = -(1 + \varepsilon)\,dt^2 + a^2 \frac{dR^2}{1 - KR^2} + a^2 R^2 d\theta^2 + a^2 R^2 \sin^2\theta d\varphi^2. \tag{15}$$

The term $(1 + \gamma)$ in Inequality (7) takes a value of 2 for a particle at rest. After the calculation of the geodesic scalar and the Christoffel connector as described in [38], one obtains Inequality (7) as:

$$|dt^2| \geq \frac{2\ell_p^2}{|G\,P^0\,\dot{\varepsilon}|}, \dot{\varepsilon} \equiv \frac{d\varepsilon}{dt}. \tag{16}$$

This expression can be re-written according to the classical energy–time uncertainty:

$$|P^0 d\varepsilon\,dt\,| \geq \hbar, P^0 d\varepsilon \equiv dE. \tag{17}$$

where $P^0 d\varepsilon$ corresponds to energy variations of the particle in the geodesic. Therefore, $P^0\,\dot{\varepsilon}$ corresponds to time-dependent energy fluctuations of the particle in the geodesic:

$$P^0\,\dot{\varepsilon} = \frac{P^0\,d\varepsilon}{dt} = \frac{dE}{dt} \equiv \dot{E}. \tag{18}$$

where dots over variables represent derivatives of the indicated variables with the time coordinate. And after recovering c in Inequality (17), one ends in SI units of time with:

$$|dt^2| \geq \frac{2c^3 \ell_p^2}{G|\dot{E}|} \sim \frac{O(10^{-34})}{|\dot{E}|} \frac{Kg\, m^2}{s}. \tag{19}$$

*2.2. Generic Solution for the FRW Space-Time Line Element for a Moving Particle in the R Coordinate*

The FRW symmetry allows simplification of calculations by considering geodesics moving in the R coordinate without displacements in the angular coordinates. In this condition only the components of proper velocities for t and R coordinates will contribute to the calculation of the Geodesic scalar:

$$G_{geo} \equiv 2G \left| U_0 \Gamma^0_{\alpha\beta} U^\alpha U^\beta \right| + 2G \left| U_1 \Gamma^1_{\alpha\beta} U^\alpha U^\beta \right|. \tag{20}$$

Expanding this expression, one gets:

$$\begin{aligned} G_{geo} \equiv\ & 2G \left| U_0 \Gamma^0_{00} U^0 U^0 + 2U_0 \Gamma^0_{01} U^0 U^1 + U_0 \Gamma^0_{11} U^1 U^1 \right| \\ & + 2G \left| U_1 \Gamma^1_{00} U^0 U^0 + 2U_1 \Gamma^1_{10} U^1 U^0 + U_1 \Gamma^1_{11} U^1 U^1 \right|. \end{aligned} \tag{21}$$

The contributing Christoffel connectors are:

$$\Gamma^0_{00} = \frac{\dot{\varepsilon}}{2(1+\varepsilon)}\ ;\ \Gamma^0_{01} = \Gamma^1_{00} = 0\ ;\ \Gamma^1_{11} = \frac{KR}{(1-KR^2)}$$

$$\Gamma^0_{11} = \frac{a\dot{a}}{(1+\varepsilon)(1-KR^2)} + \frac{a^2 KR\dot{R}}{(1+\varepsilon)(1-KR^2)^2}\ ;\ \Gamma^1_{10} = H + \frac{KR\dot{R}}{1-KR^2}. \tag{22}$$

where dots over the R coordinate, the scale factor and the metric perturbation indicate differentiation by time coordinate. Calculating the geodesic scalar with the non-zero terms one gets:

$$\begin{aligned} G_{geo} \equiv\ & 2Gm \left| U_0 \Gamma^0_{00} U^0 U^0 + U_0 \Gamma^0_{11} U^1 U^1 \right| \\ & + 2Gm \left| 2U_1 \Gamma^1_{10} U^1 U^0 + U_1 \Gamma^1_{11} U^1 U^1 \right|. \end{aligned} \tag{23}$$

And introducing the explicit terms for the Christoffel connectors one gets:

$$\begin{aligned} G_{geo} \equiv\ & 2Gm \left| U_0 \frac{\dot{\varepsilon}}{2(1+\varepsilon)} U^0 U^0 \right. \\ & \left. + U_0 U^1 U^1 \left( \frac{a\dot{a}}{(1+\varepsilon)(1-KR^2)} + \frac{a^2 KR\dot{R}}{(1+\varepsilon)(1-KR^2)^2} \right) \right| \\ & + 2Gm \left| 2U_1 U^1 U^0 \left( H + \frac{KR\dot{R}}{1-KR^2} \right) + U_1 U^1 U^1 \left( \frac{KR}{(1-KR^2)} \right) \right|. \end{aligned} \tag{24}$$

The expression can be simplified as a function of contravariant $U^0$ and the use of some equalities:

$$U^1 = U_1 g^{11} + U_0 g^{01} = U_1 \frac{1-KR^2}{a^2},$$

$$U_0 = U^0 g_{00} = -(1+\varepsilon)\, U^0, \tag{25}$$

$$U_0 U^0 + U_1 U^1 = -1.$$

which leaves the geodesic scalar:

$$G_{geo} \equiv 2G\,m \left| (-1 - U_1 U^1) \frac{\dot{\varepsilon}}{2(1+\varepsilon)} U^0 - U^1 U_1 U^0 \left( H + \frac{KR\dot{R}}{1-KR^2} \right) \right| \\ + 2Gm \left| 2U_1 U^1 U^0 \left( H + \frac{KR\dot{R}}{1-KR^2} \right) + U_1 U^1 U^1 \left( \frac{KR}{1-KR^2} \right) \right|. \tag{26}$$

and it is equivalent to:

$$G_{geo} \equiv \left| \frac{2G\,m(1 + U_1 U^1)\dot{\varepsilon}}{2(1+\varepsilon)} U^0 + 6GmU^1 U_1 U^0 \left( H + \frac{KR\dot{R}}{1-KR^2} \right) \\ + 2GmU_1 U^1 U^1 \left( \frac{KR}{1-KR^2} \right) \right|. \tag{27}$$

To simplify the expression, a curvature-associated factor, $F$, is defined:

$$F \equiv \frac{KR}{1-KR^2}. \tag{28}$$

and the Geodesic scalar takes the final form of:

$$G_{geo} = \left| \frac{Gm(1 + U_1 U^1)\dot{\varepsilon}}{1+\varepsilon} U^0 + 6GmU^1 U_1 U^0 (H + F\dot{R}) + 2GmU_1 U^1 U^1 F \right|. \tag{29}$$

Incorporating this term into Inequality (7):

$$|d\tau^2| \geq \frac{(1+\gamma)\ell_p^2}{G \left| \frac{(1+U_1 U^1)\dot{\varepsilon}}{1+\varepsilon} P^0 + 6U^1 U_1 P^0 (H + F\dot{R}) + 2U_1 U^1 P^1 F \right|}. \tag{30}$$

A flat space-time is an accurate model for current cosmological models, which corresponds to $F$ set to zero. Additionally, the epsilon correction can be ignored in $1+\varepsilon$:

$$|d\tau^2| \geq \frac{(1+\gamma)\ell_p^2}{GP^0 |6U^1 U_1 H - U^0 U_o \dot{\varepsilon}|}. \tag{31}$$

The quadratic length for the FRW space-time line element thus depends on two factors. The first one, a function derived from the energy–time uncertainty ($E_{un}$), and the second one dependent on the expansion rate of the universe ($H_{ex}$):

$$E_{un} \equiv U^0 U_o \dot{\varepsilon} \quad , \quad H_{ex} \equiv 6U^1 U_1 H$$

$$|d\tau^2| \geq (1+\gamma) \frac{\ell_p^2}{GP^0 |H_{ex} - E_{un}|}. \tag{32}$$

*2.3. Specific Solutions for the FRW Space-Time Line Element for a Moving Particle When $E_{un}$ is the Dominant Term*

One can consider several scenarios depending on which of the functions is dominant in the denominator of Inequality (32). If $E_{un}$ is several orders of magnitude larger than $H_{ex}$, then the expression simplifies as:

$$|d\tau^2| \geq \frac{(1+\gamma)\ell_p^2}{GP^0 |-U^0 U_o \dot{\varepsilon}|}. \tag{33}$$

Considering the following equalities for a particle at non-relativistic velocities:

$$(1+\gamma) = 2, P^0 \dot{\varepsilon} \equiv \dot{E} \,;\, U^0 U_o \equiv \frac{E^2}{m^2}. \tag{34}$$

And after incorporating these equalities into Inequality (33), it takes the form:

$$|d\tau^2| \geq \frac{2m^2 \ell_p^2}{G|E^2 \dot{E}|}. \tag{35}$$

The mass term, *m*, is of the same order of energy than the total energy of particles moving with non-relativistic velocities. This will lead to further simplifications, and after re-introducing *c* one gets:

$$|d\tau^2| \geq \frac{2c^5 \ell_p^2}{G|\dot{E}|}. \tag{36}$$

The minimum allowed length for the space-time line element in the FRW metric is inverse to the rate of change of energy fluctuations of the particle in the geodesic. The quadratic space-time line element is then in SI units:

$$|d\tau^2| \sim \frac{O(10^{-18})}{|\dot{E}|} \frac{Kg\ m^4}{s^3}. \tag{37}$$

As expected, Inequality (19) is recovered but in units of length.

*2.4. Specific Solutions for the FRW Space-Time Line Element for a Moving Particle When $H_{ex}$ is the Dominant Term*

Inequality (37) is derived directly by applying the uncertainty principle over the geodesic equation. As such, the inequality would not be valid in the absence of time-energy quantum fluctuations. Nevertheless, one can test the scenario in which there are not quantum energy fluctuations for large objects moving at non-relativistic velocities. Then the following equalities are fulfilled:

$$|\dot{\varepsilon}| \sim 0\ ;\ (1 + \gamma) = 2. \tag{38}$$

This condition makes the $H_{ex}$ term dominant in Inequality (32):

$$|d\tau^2| \geq \frac{\ell_p^2}{3GP^0|U^1 U_1 H|}. \tag{39}$$

To simplify this expression, the mass, m, can be incorporated into one of the radial proper velocities to convert it to proper momentum, $P^1$:

$$|d\tau^2| \geq \frac{m\ \ell_p^2}{3GP^0|P^1 U_1 H|}. \tag{40}$$

The mass-energy term, m, is of the same order than $P^0$, leading to their simplification:

$$|d\tau^2| \geq \frac{\ell_p^2}{3G|P^1 U_1 H|}. \tag{41}$$

And recovering c into the equation, in SI units:

$$|d\tau^2| \geq \frac{c^5 \ell_p^2}{3G|P^1 U_1 H|} \sim \frac{O(10^{-18})}{H\ E_k} \frac{Kg\ m^4}{s^3}. \tag{42}$$

$P^1 U_1$ is a term proportional to proper kinetic energy. The minimum quadratic length for the space-time line element in units of time, which is of the order of $10^{-18}$, is modified by the kinetic energy of the particle and the inverse of the expansion rate of the Universe. The current small value for the Hubble factor makes the space-time line element unrealistically large. In addition, when the kinetic energy of the particle is 0, Inequality (42) is undefined. Hence, it is confirmed that Inequality (37) is not valid in the absence of quantum energy fluctuations.

**3. Discussion**

In this paper, a covariant form of the uncertainty principle [38] is applied as a constraint over FRW geometries to obtain expressions for the quadratic space-time line elements in a homogeneous, isometric Universe which expands according to Hubble´s law. The imposition of GeUP introduces a quantization condition of the momentum-position phase space on classical GR geodesics. This mathematical constraint leads to quadratic space-time line elements proportional to Planck length squared. This is in agreement with most quantum gravity theories. It has to be remarked that this mathematical constraint does not truly constitute a quantum gravity theory per se. It relies on specific solutions to classical GR equations, without coupling an external mass field, unlike semi-classical formulations of quantum gravity [22]. Explicit quantization of space-time is not introduced as well such as lattice quantization. However, the results from this mathematical constraint provide some theoretical constraints to space-time line elements in specific GR solutions compatible with quantum gravity phenomena.

The experimental measurement of length scales for space-time quantization is critical to set up proper mathematical constraints for quantum gravity, and discard incompatible scenarios [13]. For example, it would help in deciding the correct lattice quantization of space in LQG, and its properties regarding the time problem and the need for privileged reference frames [5,25,42,43]. Most current quantum gravity theories, such as LQG and string theory [6,10], predict that LIV could occur by the discrete nature of space-time. Hence, the experimental testing of LIV could not only demonstrate space-time quantization, but also the scales of lengths and energies in which quantum gravity acts [18].

Putative upper limits to the constraints to LIV have been experimentally estimated by several means [14–17]. LIV is predicted by quantum gravity theories to affect energy and helicity-dependent photon propagation velocities, which could be measured when accumulated over astronomical distances. Thus, by measuring deviations of gamma ray burst photons (GRB 041219A), an upper limit of 1.1 $10^{-14}$ on the vacuum birefringence effect was estimated [16]. This constraint on LIV would translate into "spatial volume units" in the order of $10^{-42}$ cubic meters or less. Some recent studies are providing LIV violations experimentally at different energy orders, while other studies establish very stringent constraints for LIV, or even fail to detect it [13,15,16,44,45].

The constraints to the space-time line elements calculated in this paper are strictly obtained using a relativistic covariant formulation of the uncertainty principle in momentum and position 4-vectors. To obtain such constraints, GeUP [38] is applied to the FRW metric as an approximation to current cosmological models [40]. More specifically, a flat geometry condition was applied to the solution as it agrees with current observations [41]. As expected, the length for the space-time line element largely depends on energy–time uncertainty, and with a minor contribution by Hubble´s function. This last term would only be predominant if energy quantum fluctuations are negligible compared to our current H value ($H^0$).

Hence, the calculated lengths in this paper are proportional to $\ell_p^2$ in agreement with those from other quantum gravity theories. For example, the quanta of area and volume in LQG are proportional to $\ell_p^2$ and $\ell_p^3$ [7,8,12], and $\ell_p$ constitutes a natural unit in string theories and doubly special relativity [10,11,46,47]. It is reasonable to think that fundamental "volume blocks" on Planck length scales constitutes the natural quantization of space-time. Nevertheless, it has to be remarked that so far it is yet unclear whether LIV exists or that it can be tested on the current experimental energy scales.

**Author Contributions:** Conceptualization, D.E. and G.K.; methodology, D.E; resources, D.E. and G.K. All authors have read and agreed to the published version of the manuscript.

**Funding:** This research received no external funding. D.E is funded by a Miguel Servet Fellowship (ISCIII, Spain, Ref CP12/03114).

**Institutional Review Board Statement: Not applicable**

**Informed Consent Statement: Not applicable**

**Data Availability Statement:** Not applicable.

**Acknowledgments:** The authors thank Deri Blanco for her administrative help and support.